# Subjective physics

Romain Brette[1,2]

[1]Laboratoire Psychologie de la Perception, CNRS and Université Paris Descartes, 45, rue des Saints Pères, 75006 Paris, France
[2]Equipe Audition, Département d'Etudes Cognitives, Ecole Normale Supérieure, 29, rue d'Ulm, 75005 Paris, France

Corresponding author: Romain Brette, Equipe Audition, DEC, Ecole Normale Supérieure, 29, rue d'Ulm, 75005 Paris, France. Email : romain.brette@ens.fr

**Abstract**

Imagine a naive organism who does not know anything about the world. It can capture signals through its sensors and it can make actions. What kind of knowledge about the world is accessible to the organism? This situation is analog to that of a physicist trying to understand the world through observations and experiments. In the same way as physics describes the laws of the world obtained in this way by the scientist, I propose to name subjective physics the description of the laws that govern sensory signals and their relationships with actions, as observed from the perspective of the perceptual system of the organism. In this text, I present the main concepts of subjective physics, illustrated with concrete examples.

*"If we are capable of knowing what is where in the world, our brains must somehow be capable of representing this information."* (Marr 1982)

David Marr, one of the most influential figures in computational neuroscience, proposed that perceptual systems can be analyzed at three levels:

1) The computational level: what does the system do?

2) The algorithmic/representational level: how does it do it?

3) The physical level: how is it physically realized?

For example, consider the task of localizing a sound source for an animal with two ears (Fig. 1). The computational level would be: to localize the direction of a sound source, relative to the animal's head (Fig. 1A). When the source is on the left, the sound arrives first at the left ear, then slightly later at the right ear. So the algorithmic level could be: to calculate the delay applied to the left signal that makes it maximally similar to the right signal (the interaural time difference or ITD), and then calculate the direction that is compatible with that delay (Fig. 1B). In practice, this delay can be calculated by computing the cross-correlation function of the two signals and finding the time lag at which the function peaks. There are intermediate representations, the cross-correlation function and the ITD. Then the physical level could be the Jeffress model (Jeffress 1948; Joris, Smith, and Yin 1998): monaural neurons from the two ears project to an array of binaural neurons with various conduction delays, and each binaural neuron detects coincidences between spikes arriving from the two sides; the binaural neuron spikes when the ITD matches the difference in conduction delays from the two sides (Fig. 1C). It can be seen that, under some conditions, the array of binaural neurons implements the calculation of the cross-correlation function.

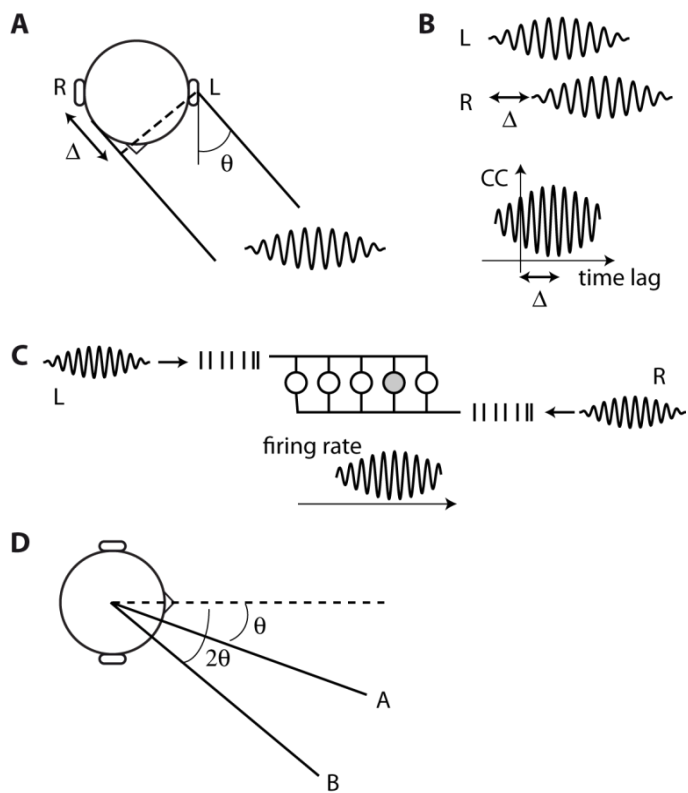

**Figure 1**. Three levels of analysis: the sound localization example. A. Computational level: an animal with two ears receives acoustical waves from a sound source at direction θ, which it must estimate. B. Algorithmic/representational level: the wave arrives at the right ear with a delay Δ, which can be extracted from the peak of the cross-correlation function. C. Physical level (Jeffress model): signals are transduced to spike trains, then transmitted with various delays to binaural neurons responding to coincidences. The firing rate represents the cross-correlation. D. Source B is at angle twice that of the angle of source A relative to the animal.

In this example, the three levels are essentially independent of each other, and this is indeed what Marr claimed when he described this methodological subdivision. But this view is not universally shared. For example, in some spike-based theories, physical instantiation is constitutive of both representations and algorithms (Deneve 2008; Brette 2012), and therefore levels 2 and 3 are not independent. More importantly for the present essay, as the quote above suggests, Marr considered that the computational level (level 1) can be defined independently of any other level, a view that Thompson et al. called "computational objectivism" (Thompson, Palacios, and Varela 1992): the function of a perceptual system is to extract objective properties of the world ("what is where in the world"), as one could describe with the laws of physics.

But on close examination of the example above, it appears that it is not so clear what is meant exactly by "what is where in the world". I will leave aside the question of "what", and focus on "where". First of all, it is obvious that the animal cannot know the absolute location of the sound source, i.e., its geographic coordinates, but only the source's direction relative to the animal. Thus, the information that an observer can obtain about the world cannot be entirely independent of itself. This is a trivial point: of course, the direction of the sound source is to be defined in ego-centric coordinates. By coordinates, we mean the angle of the sound source relative to the frontal axis. But do we mean that the animal actually calculates a number of degrees, or radians? Certainly this is absurd: these physical units are arbitrary, and in any case inaccessible to the animal. So what we mean by "angle" is not the absolute value of an angle, which has no meaning in itself, but its value relative to other angles. But then what does it mean for the observer that the angle of a sound source is twice the angle of another sound source? Quite simply: it means that if the animal turns so that it faces the first sound source, then another identical turn will make it face the second sound source (Fig. 1D). So it turns out that the very definition of spatial location is implicitly related to the potential movements of the observer. This remark was made by Poincaré at the beginning of the twentieth century, discussing the relativity of space (Poincaré 1968). In psychology, the idea that perception is tightly interlinked with the movements of the perceiver has been developed in particular by Gibson (Gibson 1986) and by O'Regan (O'Regan and Noë 2001). It is also related to embodied theories of cognition, in philosophy of perception (in particular Merleau-Ponty (Merleau-Ponty 2002)) and in robotics (for example Brooks (Rodney A. Brooks 1991)).

Thus the claim that the function of a perceptual system is to extract objective properties of the world cannot be taken literally: in fact, the computational level and the physical level are not entirely independent. But if the computational level (what is to be perceived) cannot be described in terms of objective properties of the world, then how can it be described? The aim of this essay is to define the computational level of perceptual systems, from the perspective of a perceiver embedded in its environment. In the computational-objectivist view, what is to be perceived is the properties of the world, as described by physics. Since I consider the computational level from the perspective of the perceiver, I propose that what is to be perceived is the "subjective physics" of the world – I will clarify the notion in the remainder of this text.

The following discussion of subjective physics is largely inspired by the theories of Gibson (Gibson 1986) (in particular the notion of ecological optics) and O'Regan (O'Regan and Noë 2001). It is relevant for computational neuroscience, but also for robotics, psychology and philosophy of perception.

## 1. Perceptual knowledge

*"Ever since Descartes, psychology has been held back by the doctrine that what we have to perceive is the "physical world" that is described by physics. I am suggesting that what we have to perceive and cope with is the world considered as the "environment"".* (Gibson 1972)

1.1 What is knowledge?

The goal of this section is to clarify what might be meant by "knowing what is where in the world" for a perceiver (e.g. an animal or a robot). As I mentioned above, the meaning of the term "where" is already not so obvious. The term "world" also needs clarification, as shown by the above quote from Gibson. But probably the most problematic term in this phrase is "knowing". What is meant exactly by knowledge? Let us take again the example of sound localization. From any two signals, it is possible to apply a set of operations to extract the time lag at which the cross-correlation function peaks, and then possibly to map this lag to an angle, according to a formula obtained from physics (e.g. the Woodworth formula). It could be said that by this process, some information about the source is extracted from the signals. But in fact, strictly speaking, no information is produced by this process, at least in terms of Shannon's information theory: any operation applied to a set of signals can only reduce the amount of information. At the end of the process, we simply have another number, directly derived from the initial numbers. Yet we feel that this process creates some knowledge about the world ("where" the source is). If creating knowledge is not creating information, then what is it?

Following Karl Popper in philosophy of science (Popper 1959), I suggest that what qualifies as "knowledge" is a universal statement about future observations, a *law*. Unlike a number, a law may be confirmed or falsified by future observations. Imagine that the perceiver in Figure 1 is a robot that cannot move (no motors). The kind of statements it can make upon capturing the sound waves produced by a source is: the sound wave in the right ear is the same as the sound wave in the left ear delayed by a specific amount $\Delta$. This is a law that seems to be obeyed by the two sound waves, and it could be invalidated by future observations, for example if the source moves. It is knowledge in the sense that it has a predictive value about future sensory inputs. However, there is nothing intrinsically spatial in such knowledge. If the animal can move, there can be additional knowledge, such as: if I make such movement, I will observe that the two sound waves are identical (corresponding to when the source in the front). In this case, such knowledge is about the expected outcome of actions, given the sensory observations.

Thus knowledge can be defined as universal statements or laws. I suggest calling *weak knowledge* the kind of knowledge that has a predictive value about future sensory inputs, and *strong knowledge* the kind that applies to the effect of actions. If perception is for action, then certainly the most important kind of knowledge is strong knowledge.

Note that I am not making any particular claim here about how knowledge manifests itself in the perceiver, in particular about the controversial notion of "mental representations". I simply wish to give an operational definition of knowledge, independent of the nature of the cognitive system that creates and handles that knowledge, in the same way as physics can be defined independently of the scientists that produce it.

1.2. The nature of knowledge for a perceiver

Physics qualifies as the kind of knowledge defined above. It describes the laws of nature, some of which may be relevant for perceivers, such as mechanics (walking and grasping things), optics (seeing), acoustics (hearing). For robots and animals to act in the world requires an implicit knowledge of these laws. In physics, these laws of nature are described in external terms, for example mass, tension, waves and atoms, concepts that cannot be directly grasped by organisms. Therefore organisms cannot literally understand these laws given as such. Rather, the laws that are available to them are those that govern the sensory signals they capture. For example: what happens when a limb is moved in a particular way, or how the visual field changes when the eye moves. I propose the terms "subjective physics" to describe the laws of nature from the perspective of an organism.

To contrast subjective physics with physics, I will again appeal to Popper. To distinguish science from metaphysics, Popper proposed that a scientific statement is one that can potentially be falsified by an observation, whereas a metaphysical statement is a statement that cannot be falsified. For example, the statement "all penguins are black" is scientific, because I could imagine that one day I see a white penguin. On the other hand, the statement "there is a God" is metaphysical, because there is no way I can check. Closer to the matter of this essay, the statement "the world is actually five-dimensional but we live in a three-dimensional subspace" is also metaphysical because independently of whether it is true or not, we have no way to confirm it or falsify it with our senses.

In the same way, knowledge about the world that qualifies as non-metaphysical for a given perceiver depends on the senses it possesses and the actions it can make. For example, imagine that the perceiver in Figure 1 is a fixed robot, with no ability to move. In this case, all it can grasp is the delay between the two sound waves. From this delay, the angle could be inferred using the Woodworth formula for example, so let us consider the following statement: "the sound source is at angle x". From the point of view of an external observer, this is a scientific statement, because she can measure that angle with a tool. However for the perceiver, this is a metaphysical statement because the statement cannot be falsified. Now consider that the robot can turn its head. Then the statement "the sound source is at angle x" is still a metaphysical statement. A non-metaphysical statement would be "the sound source is at such a location that if I make movement x, the two sound waves will be identical". That is, the location is defined only in terms that are accessible to the perceiver.

Thus, I define subjective physics for a perceiver as the laws that govern the sensory inputs and the effect of actions on them, as they are implied by physics, but only including those laws that are non-metaphysical for that specific perceiver.

1.3. Subjective physics and Gibson's ecological optics

What I just described as subjective physics is very close to what Gibson described as "ecological optics" for vision. The word "ecological" refers to the relation between a specific organism and its environment, and therefore is fully relevant to this matter. But there are a few reasons why I favored the terms "subjective physics" over "ecological physics". The first reason is that I wanted to avoid the confusion with environmental physics or ecology. The second reason is that I want to make clear that subjective physics is not a psychological theory. It is relevant to psychological theories, but it does not rely on specific assumptions about the cognitive abilities of the perceiver, or about what perception is. It is only meant to describe what is intrinsically available to a perceiver, given a specific set of sensors and actuators. Thus, the term "subjective" should only be understood as "from the perceiver's perspective". Finally, a third reason is that Gibson described the laws of ecological optics in terms of the structure of the visual field (the "optic array"), independently of the fact that light is received by sensors. But this fact is highly significant, because the activity of these sensors is only indirectly related to the optic array. Notably, in the eye there is a blind spot, there are inhomogeneities in spatial sampling but also in color sampling. These facts and their significance are taken into account by O'Regan's sensorimotor theory of perception (O'Regan and Noë 2001; O'Regan 2011). This extensive quote explains the nature of the problem very well with an experiment of thought:

*"Imagine a team of engineers operating a remote-controlled underwater vessel exploring the remains of the Titanic, and imagine a villainous aquatic monster that has interfered with the*

*control cable by mixing up the connections to and from the underwater cameras, sonar equipment, robot arms, actuators, and sensors. What appears on the many screens, lights, and dials, no longer makes any sense, and the actuators no longer have their usual functions. What can the engineers do to save the situation? By observing the structure of the changes on the control panel that occur when they press various buttons and levers, the engineers should be able to deduce which buttons control which kind of motion of the vehicle, and which lights correspond to information deriving from the sensors mounted outside the vessel, which indicators correspond to sensors on the vessel's tentacles, and so on."*

To rephrase it, in addition to Gibson's ecological optics, subjective physics acknowledges that the specific relation between physical inputs (light) and sensor activity (input signal to the perceptual system) qualifies as metaphysical knowledge for the perceiver – i.e., it cannot be known from the sensory inputs alone.

Thus subjective physics has been used before as a component of psychological and philosophical theories of perception, but it has not been developed for itself. The goal of this text is to provide definitions and relevant concepts for subjective physics. I suggest that subjective physics is particularly relevant to computational neuroscience, but it is also relevant to psychology, neuroscience, robotics, and philosophy of perception. It is an attempt to redefine what Marr called the "computational level" of perceptual systems, without recourse to metaphysical knowledge about the world. It provides an alternative to the more traditional approach, in which the computational level is presented as an inverse problem to be solved, i.e., recovering objective properties of the world using metaphysical knowledge about the relation between objects and sensors.

But what can be known without metaphysical knowledge about the world, just from a set of sensors and actuators?

## 2. A detailed example

### 2.1. Subjective physics of a hearing robot

To clarify the problem, I will discuss again the simple example of the hearing robot, but now in more detail (Fig. 2). On top of the robot's head, there are two antennae, with one microphone mounted on each one, at the same height. The two microphones are close to each other. In the world, there are sound sources, which produce sounds repeatedly, and they lie on the floor. There is only one source present in the world at a time, and it is present for a long time. What can the robot know about the world, without metaphysical knowledge?

First of all, when a source produces a sound, two sound waves are captured by the two mics, and these two sound waves have a very special property, which is that they are delayed versions of each other (Fig. 2A). Strictly speaking, the robot cannot examine the sound waves themselves but rather the electrical signals produced by the mics. However, for now, I will assume that this conversion is identical for the two mics, so that it will not affect the following discussion.

To notice this law, or "invariant structure" in Gibsonian terminology, may require quite advanced cognitive abilities, but the present discussion focuses on what the laws of subjective physics are (the computational level), not on how they are extracted (the algorithmic level). The important point to notice here is that this law can be phrased independently of any metaphysical knowledge about sound, or even about the fact the signals are acoustical. It is simply that two signals are delayed copies of each other.

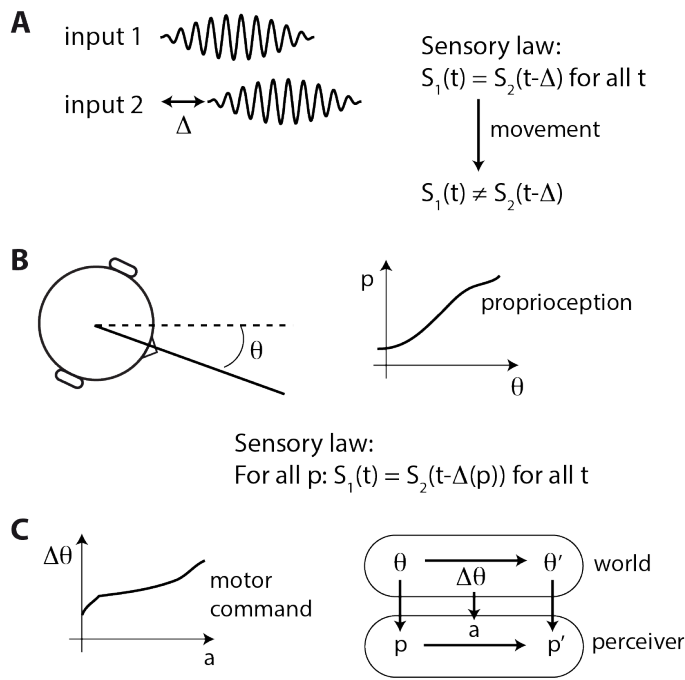

**Figure 2**. Subjective physics of a binaural robot. A, When a source produces a sound, the robot captures two sensory signals $S_1$ and $S_2$. The robot can notice that these signals follow a particular law: $S_1(t) = S_2(t-\Delta)$ for all t. The law is spatial because it is falsified when a movement of the head is produced. B, The robot receives a proprioceptive signal p related to the head's angle (possibly in a nonlinear way). There is a relationship between p and a sensory law followed by the auditory signals, which defines the sound location (note that the function $\Delta(p)$ depends on source location). C, A motor command *a* produces a rotation of the head. Each action *a* produces a change in the proprioceptive signal p. The relationship to physical quantities is unknown (world) but the structure is the same (group action of motor commands on proprioceptive signals).

When the same sound source produces another sound, the same property is true, with the same delay. But when another sound source produces a sound, the property holds but with another delay. Although we (external observers) know that the value of this delay is related to the direction of the sound source (angle θ), at this stage the robot has no way to know it – it would require metaphysical knowledge. To see this clearly, consider again the thought experiment proposed by O'Regan and Noë (2001): imagine the wires from the two mics have been mixed up and you do not know which one belongs to the left or to the right mic. Then given the delay between the two signals, it is not possible to tell whether the sound source was on the left hemifield or on the right hemifield. Thus there cannot be any spatial knowledge at this stage.

We now consider the fact that the robot can rotate its head. When the head rotates, the observed delay changes. Thus the sensory property that was picked up, the fact that the two sound waves are delayed copies of each other with a specific delay, is falsified when a movement is produced.

In this sense it can be said that the property is spatial. In addition, when the head rotates, the two sound waves are still delayed versions of each other, but the delay changes in a systematic way with the movement: there is a lawful relation between the head angle and the delay (Fig. 2B). At this point we must be careful: if there is no metaphysical knowledge, then there is in fact no such thing as "head angle". Rather, the robot has access to either proprioceptive signals related to that angle, and/or to motor commands. For now, I will assume the former: the robot has access to another sensory input that is related to head angle by a possibly nonlinear relation. Thus, from the perspective of the robot, there is a lawful relation between proprioceptive signal and interaural delay. There is a one-to-one correspondence between this relation and the angle of the sound source, and so in this sense it may be said that the direction of the sound source can be picked up by the robot, in the form of the higher-order sensory relationship between head proprioception and interaural delay (a lower-order sensory relationship). Note that the villainous monster is not a problem at all here: the same structure exists if the mics are inverted.

Several remarks are in order. The relation that defines the direction of the sound source is not defined directly on the sensory signals. It is defined as a relation between one sensory signal (proprioception) and a set of relations on sensory signals (acoustical inputs). It is a relation between a value and a relation, and therefore it can be described as a higher-order relation. The second point is that the notion of space is very restricted here. In particular, the direction can be known but not the distance. But even then, the notion of direction is very weak. There is a notion of topology, that is, that directions can be arranged on a circle and not on an infinite line. But there is no metric structure: it cannot be known that an angle is twice another angle for example. Mathematically, the group structure of angles is not part of the subjective physics of the robot's world.

Consider now that, in addition to proprioceptive signals, the robot also has access to its motor commands – what is called the "efferent copy" in neuroscience (Fig. 2C). I assume that these commands take the form of rotation commands. A rotation command is an action that specifies a rotation with a value that relates to the angle of the rotation, but again the relation between the value and the actual physical angle is unknown. Now when the robot performs an action $a$, in the form of a specific rotation, the proprioceptive signal changes in a specific way, from p to p'. Thus action can be considered literally in the mathematical sense as a mapping from $(a,p)$ to $p'$, which one would write $a.p = p'$. It can be seen that this action has the algebraic structure of a group action. Indeed, for every action there is another action that brings the structure back to its original state (corresponding to the inverse rotation), which is then called the inverse action; the combination of two actions is another action. The group is also commutative because the order of actions has no effect on the end result. Note how this group structure arises even though rotations are not specified as angles or even as quantities linearly related to angles. The subjective physics of the robot now includes a much richer notion of space, where actions are isomorphic to the group of rotations. In particular, it now makes sense to say that an angle is twice another angle.

2.2. Inference vs. pick-up

A valid objection to the above remarks is that observing the sensorimotor relation that defines source direction requires making movements, and so it cannot be picked up on the first presentation of the sound. Wouldn't that mean that a single short sound cannot produce spatial perception? This is where inference becomes important and I must depart from Gibson (Fig. 3).

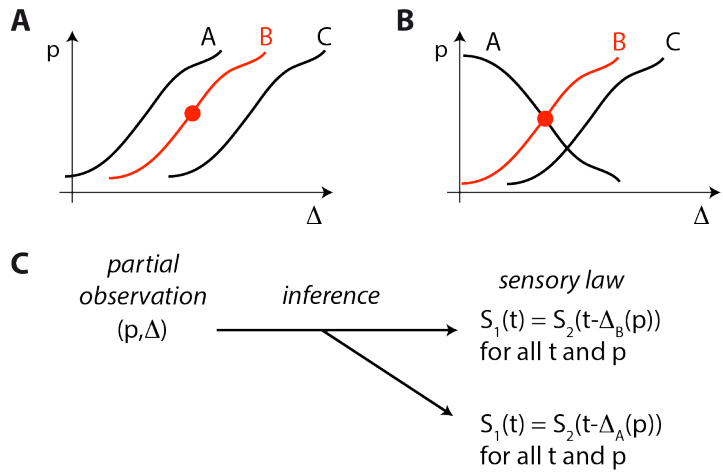

Figure 3. Inference and pick-up. A, For each source location (A, B, C), there is a relationship between delay Δ and proprioceptive signal p, which defines that location. This relationship can be picked up by producing movements if the sound is continuous (red curve), but if the sound is transient then only a partial observation (p, Δ) is available (red dot). In this case source location (the full curve) is *inferred* if the partial observation is consistent with some previously observed relationship. B, Inference can be ambiguous if several relationships are consistent with the observation (for example a sound coming from the front or from the back). C, In this framework, inference means hypothesizing a particular sensory law from partial observations, which can be an ambiguous process.

When a continuous sound is presented at a given location, there is a lawful relation between head position p and interaural delay Δ that can be "picked up" by producing movements. The terms "picked up", based on Gibson terminology, mean that no memory is required to produce this knowledge, it simply derives from the present observations. When only a single observation (p, Δ) is available, such pick-up is not possible (Fig. 3A). However, based on previously observed laws, one may notice that the observation is consistent with one of these laws, in which case it is *inferred* that the future observations (p, Δ) should follow the identified law – for example if the source produces sound again. Since the observation may be consistent with several laws, inference may come with some degree of ambiguity (Fig. 3B). For example, Δ = 0 for sources in the front and in the back, but the delay changes in opposite directions with head angle (sources A and B in Fig. 3B). This is inference and not pick-up because it relies on previously acquired knowledge (Fig. 3C). But note that unlike in traditional approaches, inference does not rely on metaphysical knowledge (a priori assumptions about what should be inferred). Again, how this inference is made by the perceptual system is an important and difficult question, but one that concerns the algorithmic level, not the computational level. Here I simply point out that there are cases that require inference, and that what is to be inferred is related to previously picked-up laws.

Let us summarize our findings so far. There are two properties that can be picked up. There is a low-order sensory property, involving no action, the property that the two monaural signals are delayed copies of each other, with a specific delay. It is spatial in the sense that it is affected by movements. There is then a higher-order property, the relation between head position and the low-order interaural property. Both types of properties are sensory, but additionally rotation commands form a group action on head position. All this structure can be "picked-up", i.e., it can be discovered without relying on preexisting knowledge – this is presumably what Gibson meant when he insisted that perception is "direct". Then through learning, relations can be inferred on the basis of partial observations, using the previously picked-up knowledge – this is arguably not direct perception. The example of the transient sound (e.g. a hand clap) is particularly interesting: in this case, one can pick-up the property that the two signals differ by a particular delay, but one must infer the relation between future actions and the change in delay.

We can say more about this example. In particular, we note that the inferential process is indeed ambiguous: the same interaural delay can correspond to two different directions that are symmetrical with respect to the interaural axis, a "front-back confusion" (Fig. 3B). This is really an ambiguity for the inferential process, not an ambiguity in the laws of subjective physics, as is the case for distance. Indeed there is no front-back confusion in the direction as defined by the sensorimotor relationship. The confusion only exists in the mapping from interaural delay to direction. Thus, the spatial property to be perceived in the robot's world is direction defined on the entire circle, and the inferential process has a front-back ambiguity, which can be resolved by movements.

A final remark regarding inference and pick-up: strictly speaking, given that there can only be a finite number of (possibly unreliable) observations, all knowledge is inferential. For the matter of describing subjective physics, this is not a critical issue. It should be understood that subjective physics describes the underlying laws that can potentially be discovered by the perceiver, in the same way as physics describes the laws of the world even though human observation is limited. The term "pick-up" refers to the ideal limit of unrestrained observation.

2.3. Cognitive abilities

In this very simple example, it appeared that the information available to the robot without metaphysical knowledge is surprisingly rich. Indeed, the robot can have a sense of direction, isomorphic to a circle, along with its topological structure. By describing this information in a systematic way, we have also uncovered that to obtain and use that information, the robot must display a number of cognitive abilities, such as:

- the "pick-up" of low- and high-order sensory relations
- producing movements
- long term memory (for inference)
- inference

Thus the aims of subjective physics are two-fold: 1) to describe "what is out there", that is, what information is available to the perceiver within an environment, given a set of sensors and actuators, 2) to uncover the cognitive abilities that are necessary for an organism to discover this information. Thus subjective physics hints at the algorithmic level, although it does not describe it. I want to stress again that the notion of "information" about the world must not be understood in the sense of Shannon, because Shannon's information is unstructured and its interpretation requires metaphysical knowledge (the "code"). The information or knowledge we are talking about takes the form of laws, which can be arbitrarily complex and structured – exactly like the laws of physics.

## 3. Definitions and basic principles

The goal of this section is to define the core concepts of subjective physics. First of all, I define "subjective physics" as the field of study that analyzes the structure of the sensory and sensorimotor relationships that are available to an organism embedded in an environment, given a set of sensors and actuators, without "metaphysical knowledge" about the world (which I explain in more detail below). I shall call this structure the "subjective structure of the world". It is directly related to what Gibson and followers described as "ecological optics" (Gibson 1986) and "ecological acoustics" (Gaver 1993). It is not at all meant to be a psychological field, even if there are relationships with psychological theories. No assumptions are made about the cognitive abilities of the organism. The aims are only to describe what information is *available* to the organism, not what perception is for that organism. The mode of enquiry is agnostic about the psychological question, and in fact it is meant to apply to living organisms as well as to robots. The organism (animal or robot) will be referred to as the "perceptual system".

The aim of subjective physics is to analyze in the highest possible detail the subjective structure of the world, and in particular under what form the information is available. As we have seen in the example, it is also expected that this descriptive process uncovers some of the cognitive abilities that are necessary for the perceptual system to discover this information.

The analysis is specific to a set of sensors and possible actions on the world. I will call this set the "interface" of the perceptual system. What is meant exactly by "sensors" and "actions"? A sensor is simply something that is modified by the world, and whose modification can be picked up by the perceptual system. For example, acoustical waves make the membrane of a microphone vibrate, and this vibration can be picked up by the robot. An action is something that the perceptual system can do, which produces modifications in the world. In turn, these modifications can affect sensors.

There are a few subtleties in these two concepts that I will discuss in section 3.2. But to uncover these subtleties, I first need to introduce a methodology that I shall call "ecological reduction" in reference to the concept of "phenomenological reduction" in philosophy.

### 3.1. Ecological reduction

Subjective physics aims at describing the subjective structure of the world without recourse to metaphysical knowledge of the world. I introduced the terms "metaphysical knowledge" in reference to Popper's demarcation criterion in philosophy of science. It refers to statements about the world that cannot be falsified or corroborated given a specific interface with the world (sensors and actions). Thus what is considered as metaphysical is always with respect to a specific interface.

There is an interesting parallel to be made with phenomenology in philosophy. The key point of subjective physics is to get rid of any preconception about the nature of the external world. This is not to say that such preconception, such as the Euclidian structure of space, is not real, but we wish to suspend this belief while describing the subjective structure of the world, so as to focus on what is really intrinsic to that structure. This attitude is in fact strikingly analogous to the aims of phenomenology, a field of philosophy initially introduced by Edmund Husserl at the beginning of the 20th century, and later developed by a number of philosophers, such as Heidegger, Merleau-Ponty and Sartre. Phenomenology is the philosophical study of the structures of subjective experience and consciousness. This study relies on Husserl's methodology of "phenomenological reduction" (also called "bracketing" or "epoché", which means suspension in Greek), which is the suspension of any a priori judgment about the world. This is not to say that the object of consciousness is denied any objective existence, or to refute that it has objective properties, but simply to refrain from using this knowledge in the analysis of the phenomenon. For example, a phenomenological analysis of color would not allow physicalist statements such as: "red is the conscious experience of light with wavelength 650 nm".

Similarly, subjective physics relies on the methodology of "ecological reduction", that is, the suspension of any prior knowledge about the world on behalf of the perceptual system, in the analysis of the subjective structure of the world. This suspension may extend to the nature or properties of the sensors and the specification of actions (see section 3.2).

An associated concept in phenomenology is "eidetic variation" or "eidetic reduction" ("eidos" means shape or form in Greek, as in Plato's "ideal Forms"). It consists in varying the perspective on the phenomenon so as to clearly understand what constitutes its essence, what is critical to the phenomenon and what the phenomenon is invariant to. Applied to subjective physics, this means slightly varying the nature of the external world (for example the nature of the sound sources), the interface with the world (nature of sensors and actions that can be performed, for example whether the robot has proprioceptive information), or possibly the constraints on the perceptual system (for example whether it has memory). This technique allows a deeper analysis of the structure of the world, as it reveals how the structure depends on the specification of the world and of the perceptual system.

To be more concrete, O'Regan's "villainous monster" (O'Regan and Noë 2001) is an application of this reduction technique to the interface with the world: if sensors are mixed up in an unknown way, what knowledge about the world remains? By this technique, one reveals and discards metaphysical knowledge about sensors.

3.2. Sensors and actions: the interface of the perceptual system

*3.2.1. Sensors*

A sensor is something that is modified by the world, and whose modification can be picked up by the perceptual system. I gave the example of two microphones. This may give the impression that the perceptual system has access to the value of the acoustical pressure at any given moment in time, at both ears. I implicitly assumed it when I described the example. But there are in fact two presuppositions in this statement:

1) The two sensors have the same properties. I will call this assumption the "sensor homogeneity" assumption.

2) The values provided by the sensors can be interpreted as acoustical pressure, which qualifies as metaphysical knowledge.

I will start with the second point. In fact, a microphone does not exactly provide the acoustical pressure, it provides an electrical signal (Fig. 4A). The signal is a function of the acoustical pressure, but to deduce the acoustical pressure from the electrical signal requires the knowledge of this function. We may want to free ourselves from this assumption in the analysis of the subjective structure of the world. In the robot example, we can easily remove this assumption, because what matters is only that the two sensor signals are delayed copies of each other. Making the sensors nonlinear would also leave the subjective structure of the world unchanged (note how we used eidetic variation to analyze the determinants of this structure).

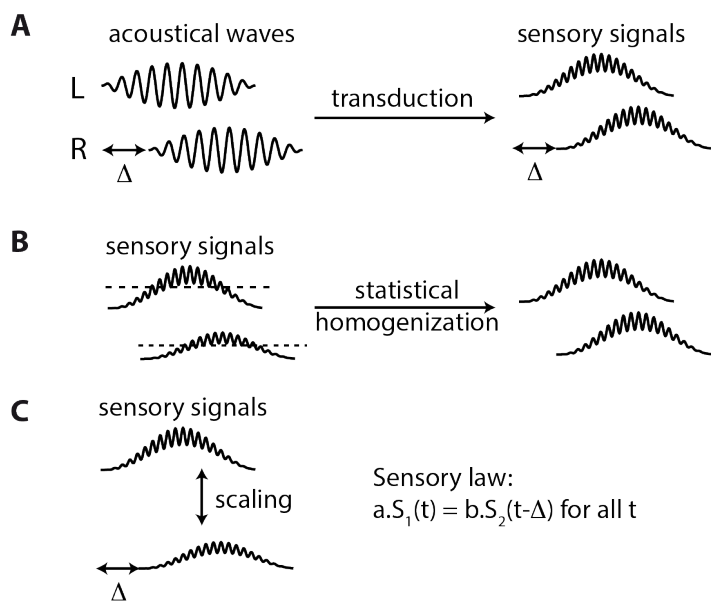

**Figure 4**. Sensory signals. A, Sensory signals are determined by acoustical inputs through an unknown and possibly nonlinear transformation, but here the structure (delay) is preserved. B, If sensors are not homogeneous, the structure may be changed. One way to solve this problem is statistical homogenization: each signal is normalized with respect to some statistics (e.g. average, dashed line). C, Another way is to allow more general sensory laws: here the two signals are related by a temporal shift, which depends on the source, and a scaling transformation, which is universal (depending on the sensors only).

The sensor homogeneity assumption is more difficult to remove. In the example, it is not a problem at all that the sensor signal is an indeterminate function of acoustical pressure. But if the two sensor signals are differently related to acoustical pressure in the two microphones, then they would not be delayed copies of each other anymore. It is easy to imagine such a case: there is some tolerance in the properties of electrical components, and so the two signals could be similar and proportional to each other but no exactly identical. It is even easier to imagine in an animal. Let alone the variability in receptor properties, we may simply consider the eye: given the blind spot due to the optic nerve, the blood vessels, the optical aberrations and the fact that light must go through several layers of cells before reaching the receptors, it is clear that two photoreceptors do not exactly see the same thing up to a spatial shift – this is a critical observation in the sensorimotor theory of perception (O'Regan and Noë 2001; O'Regan 2011).

I will describe two procedures to remove the sensor homogeneity assumption. The first one is "statistical homogenization" (Fig. 4B). Suppose the two microphones produce signals proportional to acoustical pressure, but with different and unknown proportionality factors. One could decide to scale these signals by two different factors, chosen so that the two resulting signals have the same average power (on a long timescale). This procedure solves the problem of sensor inhomogeneity. But we note that it does not provide a way to recover the true acoustical pressure, it only equalizes the signals. Indeed it is impossible to know such an objective value, because only relationships can be observed. More generally, statistical homogenization consists in choosing a particular statistics and transforming the sensor signals so that they have identical long-term values for this statistics. We note that the procedure relies on long-term learning, on behalf of the perceptual system.

Another procedure is to extend the types of relationships that are observed on the sensory signals (Fig. 4C). Suppose the two acoustical signals are given by X(t) and Y(t). The presence of a sound source at a given direction is attested by the relationship X(t)=Y(t-d) for all t, for a specific delay d. But the perceptual system only has access to $S_1(t)$ and $S_2(t)$, which are scaled version of X and Y, with unknown scaling factors. Then the presence of a sound source is attested by the following relationship: there exist two numbers a and b such that $a.S_1(t)=b.S_2(t-d)$. The numbers a and b are universal in the sense that they are identical for all source directions. Thus, the procedure is to allow the discovery of more complex relationships in the sensory signals. Again, this does not provide a way to recover the original acoustical pressure. On behalf of the perceptual system, this means that it must be possible to discover complex relationships in the signals.

What we have just done illustrates the concept of "ecological reduction": we progressively removed implicit metaphysical knowledge of the world, and analyzed what structure remains. It appears that the spatial structure of the world is robust: it remains even when sensors are inhomogeneous and related to acoustics in an unknown way. On the other hand, the objective acoustical pressure or the energy of the signals cannot be known from such sensors. We also observe that dealing with sensor inhomogeneities puts an additional load on the perceptual system, in terms of cognitive abilities. Note that we could go further in this reduction, for example by not assuming that the two sensors have the same spectral response.

The previous observations suggest that the subjective structure of the world can be analyzed at different depths of ecological reduction: at the first level, we consider that sensor data is specified in external terms (acoustical pressure); at the second level, we consider that sensor data is specified in internal terms (transduced quantity); at the third level, we consider that sensors are inhomogeneous.

*3.2.2. Actions*

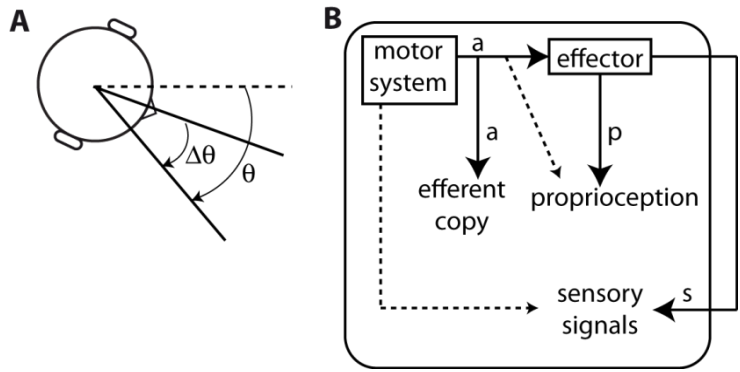

**Figure 5**. Actions and their effects on sensorimotor signals. A, A motor command for the robot may be specified as a target angle Ө, or as the angle increment ΔӨ. B, A motor command *a* is sent to an actuator (muscle or motor), which results in different types of signals available to the perceptual system: the efferent copy of the command *a*, the proprioceptive signal *p*, and the sensory signals *s* due to the interaction of the actuator with the world. In addition, motor commands may affect proprioceptive or sensory signals without interaction with the world (dashed lines), for example the efferent system in the cochlea (Guinan 2006).

An action is something that the perceptual system can do, which produces modifications in the world. In turn, these modifications can affect sensors. For example, the robot can rotate its head. This definition needs to be made more precise (Fig. 5A). Indeed: do we mean that the perceptual system issues a command that makes its head rotate *to* a particular angle (relative to the fixed body), or *by* a particular angle (relative to the current head angle)? In the first case, the command is issued in an absolute spatial frame. In the second case, the command is issued in a relative spatial frame. This has important consequences for the subjective structure of the world, and for the cognitive load on the perceptual system. Indeed, the interaural delay is relative to the head, not to the fixed body. Therefore, if the reference frame for commands is the head, then there is no way that the perceptual system can capture the absolute direction of the sound source, unless there is proprioceptive information. If the reference frame is the fixed body, then inferring the direction from a single sound presentation requires either proprioceptive information or memory of the last command. The sensorimotor structure is also different in these two cases.

Secondly, we can make the same remark about actions as about sensors. Considering that the command is issued as an angle implicitly assumes metaphysical knowledge about the world, that is, about the relationship between the command itself (an electrical signal) and the result of the command in externally defined terms (rotation angle). If we want to remove this assumption, then actions should be thought of as "levers" or switches: some command that can be triggered with an associated value, which has an unknown effect on the world – apart from what is picked up by the sensors.

Actions generate two types of signals for the perceptual system: sensory signals (including proprioception) and "efferent copy". Sensory signals result from the effect of actions on the world. The efferent copy is the copy of the command values issued by the perceptual system, considered as input signals. The sensorimotor structure is the relationship between these two types of signals. The efferent copy depends only on the commands, but the sensory signals depend on both the commands and the world, which is why the sensorimotor structure is informative about the world.

A special type of sensory input is proprioception: these are signals about the body (e.g. muscle tension, position of the head relative to the body) rather than about the external world. As far as sensorimotor structure is concerned, these can be considered as sensory signals. However, their sensorimotor structure has distinctive properties, which I will discuss later. In fact, many sensors are both proprioceptive and sensory: for example, in the case of the robot, the acoustical inputs depend both on the direction of the sound source (world) and on the position of the head (body).

A special type of action is when the action influences the sensors themselves rather than the external world (although the distinction is somewhat arbitrary). For example, the pupil in the eye can contract and dilate, which changes the amount of light coming into the retina. In the cochlea, the medial efferent system changes the way incoming sounds put the basilar membrane in motion (Guinan 2006).

*3.2.3. Relationship between sensors and actions*

A critical concept in the study of subjective physics is that action is not considered as caused by sensory input. In this sense, it is not a subfield of psychology, because the object of study is not what determines an organism to act in the way it does, but rather how potential actions modify sensory inputs. It is the opposite perspective of many studies in neuroscience, in which ones observes the effect of stimuli on the nervous system or on behavior. In subjective physics, it is in fact considered that actions are *voluntary*, in the sense that any action can potentially be taken – it is not constrained by the sensory inputs. The term "voluntary" should not be taken in a psychological sense, but rather as meaning free from any determinism.

This notion of "voluntary action" is important because it makes action very different in nature from sensory inputs. One can observe relationships between sensory inputs, but it is not possible to see one input as being caused by another input: only correlates can be observed. In contrast, causality exists in the sensorimotor structure precisely because action can be taken or not taken, independently of sensory inputs. Action solves the problem raised by David Hume, that correlation does not imply causation.

It should be clear that the subjective structure of the world depends not only on the sensors but also on the possible actions. For example, if the robot could not rotate its head, the acoustical structure would remain but it would not have a spatial character (no movement can disrupt it), and it would not be possible to define a source direction (only an interaural delay).

3.3. On the notion of information

To avoid confusions, I have tried to avoid the term "information", and replaced it with the term "structure". The term "information" may indeed be confusing when speaking of knowledge about the world, because in neuroscience it is often meant in the sense of Shannon. In this subsection, I want to make the distinction as explicit as possible to avoid misunderstandings.

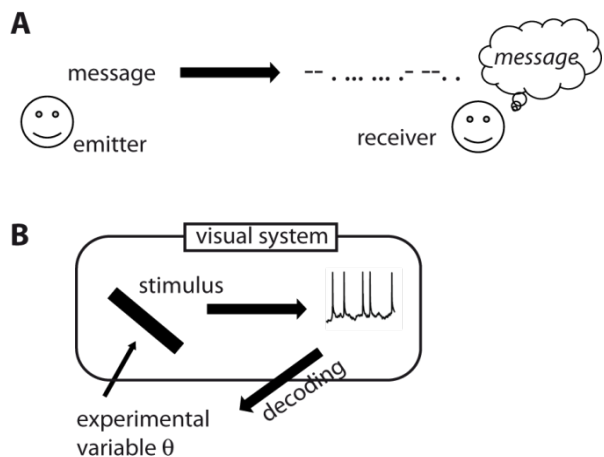

**Figure 6**. Information in the sense of Shannon. A, A communication channel consists of an emitter who wants to transmit some message to a receiver, in an altered form named "code" (here Morse code). The receiver knows the correspondence and can reconstruct the original message. B, In a neuroscientific context, the emitter is the experimenter, who presents a stimulus (oriented bar) characterized by some objective values (orientation θ). The brain receives the message in the form of neural activity, from which it infers information about the stimulus (θ). However the decoding process, i.e., the relationship between neural activity and the experimental variable, is metaphysical knowledge for this perceptual system.

Shannon's information comes from communication theory. There is an emitter who wants to transmit some message to a receiver (Fig. 6). The message is transmitted in an altered form called "code", for example Morse code, which contains "information" insofar as it can be "decoded" by the observer into the original message (Fig. 6A). The metaphor is generally carried to neuroscience in the following form: there are things in the external world that are described in some way by the experimenter, for example bars with a variable orientation, and the activity of the nervous system is seen as a "code" for this description (Fig. 6B). It may carry "information" about the orientation of the bar insofar as one can reconstruct the orientation from the neural activity.

It is important to realize how weak and specific this notion of information is. In a communication channel, the two ends agree upon a code, for example on the correspondence between letters and Morse code. For the receiving end, the fact that the message is information in the common sense of the word, i.e., knowledge about the world, relies on two things: 1) that the correspondence is known, 2) that the initial message itself makes sense for the receiver. So the notion of Shannon's information applied to a perceptual system carries with it two elements of metaphysical knowledge: the relationship between sensor signals and externally defined properties of objects in the world, and the meaning of these objects.

In the robot example, the interaural delay between microphone signals is information about the source's angle in Shannon's sense. But to infer the source angle from that delay requires knowing the relationship between the quantities, which is metaphysical knowledge (it is not included in the sensor signals). Then knowing what the angle means requires some metaphysical knowledge of Euclidean geometry.

Thus what I have called the subjective structure of the world is not at all information in Shannon's sense. It is closer to the notion of scientific knowledge: universal statements about the world, as the laws of physics, rather than code words that stand for things in the world. The distinction has two important implications. One is that knowledge about the world is highly structured, contrary to Shannon's information, because it is in the form of laws. I will discuss this point in more detail in section 4. The second implication is that such knowledge necessarily implies the notion of a sensory flow. Indeed a law cannot exist in a single observation, it can only be seen through a flow of observations. This stands in contrast with a standard statistical learning framework in which an element of information is an image (seen as a vector of pixels). Since even the most elementary law is a relationship, it cannot involve a single image.

It is important to be fully aware of this distinction with Shannon's information, because the confusion seems to subtend a number of misunderstandings, in particular about Gibson's notion of information. Gibson considered that there is intrinsic information about the world in the "invariant structure" present in the sensory flow, as explained by this quote:

*"A great many properties of the array are lawfully or regularly variant with change of observation point, and this means that in each case a property defined by the law is invariant".* (Gibson 1972)

This quote makes it clear that Gibson did not mean information in Shannon's sense but truly in the sense of a law, i.e., a universal statement about the sensory flow. The terms "invariant structure" stand for nothing else than a sensory law, as is made explicit in the quote above. It does not mean for example that some sensory signal is constant.

Subjective physics describe information about the world in the sense of structured knowledge in the form of laws, not in the sense of Shannon.

3.4. Noise, finiteness and inference

It is assumed that the perceptual system can discover the subjective structure of the world, that is, using Gibson's terminology, it can "pick-up" the sensory and sensorimotor relationships. For example, it can observe that when a source produces sound, the two sound waves at the ears are delayed copies of each other. However, one may object that inferring a law from observations, which are necessarily finite and possibly noisy, is somewhat arbitrary: there are always an infinite number of laws that are consistent with a finite number of observations.

This is a relevant objection, but it is not an objection to the existence of the law: the argument emphasizes the difficulty for the perceptual system to infer the law. In fact, this and related questions are a major theme of the philosophy of knowledge (epistemology): how can knowledge be acquired? Subjective physics does not aim at answering this question, but it is assumed that knowledge can indeed be acquired by the perceptual system. Therefore I will summarize a few relevant arguments from philosophy of knowledge.

The argument that it is not possible to infer a universal law from a finite number of observations is the skeptic criticism of inductivism. It can be addressed by different means. One is "Occam's razor": the idea that among competing hypotheses, the most parsimonious one should be preferred (a principle for which statistical learning theory gives some justification). Another way of addressing the problem was provided by Karl Popper: first one makes a hypothesis, in the form of a universal law, then the hypothesis is tested by future observations. Note that this view, falsificationism, does not actually suggest any particular method to produce the hypothesis in the first place, but for the problem at hand, it emphasizes that 1) knowledge is for predicting the future rather than for accounting for the past, 2) knowledge is useful if it has predictive power, even if it might be amended in the future. This is just a glimpse of a few relevant concepts from philosophy of knowledge. In fact, the relationship between perception and philosophy of knowledge is deep. For example, following the falsificationist account of knowledge, action could be seen as an experiment, chosen to test the current hypothesis about the subjective structure of the world (this is related to the idea of active learning (Cohn, Ghahramani, and Jordan 1996)).

To summarize this point: it may indeed be a difficult problem for a perceptual system to infer laws present in the sensorimotor signals, but these laws exist independently of the cognitive abilities of the perceptual system. The primary object of subjective physics is to describe these laws, without any particular claim about how the perceptual system could notice them. If it is assumed that a scientist can discover the laws of physics, then a perceptual system can grasp the laws of subjective physics.

## 4. Subjective structure

In this section, I will describe a few general concepts of subjective physics, taking again the example of the robot.

### 4.1. The role of action in subjective structure

As I previously argued, relationships can only be seen through a sensory flow, as opposed to a static pattern of inputs (an image). This means that something must change: either the organism moves (actions) or something in the world changes by itself. The latter can be a movement of an object or of an animal, or changes in acoustical pressure (a sound) or in illumination. An intermediate possibility is that the organism is made to move, for example if she is sitting in the passenger's seat of a car.

The sensorimotor account of perception (O'Regan and Noë 2001) emphasizes the relationship between action and the resulting sensory signals (or changes in sensory signals), and indeed I have observed that action must be involved to define any kind of spatial structure. In this section I will examine the role of action in subjective structure. At first sight, such a sensorimotor relationship may be thought of as a mapping between actions and sensory signals. However, the robot example shows that this is a little too restrictive.

Action can be involved in the structure in five different ways.

1) First of all, it can be not involved at all. For example, if a sustained note of a musical instrument is played, a periodic sound is captured at each microphone. This periodicity structure exists independently of any movement, and it is unaffected by action.

2) It can also be involved negatively, that is, structure conditioned to the absence of action. For example, when a sound is produced by a source, two acoustical waves are captured at the two microphones that are delayed copies of each other: this structure exists as long as the robot does not turn its head. If the robot moves, the structure is changed. This is what gives its spatial character to this structure, which is otherwise sensory rather than motor.

3) Action can be involved as the cause of the sensory flow. For example, when a bird flies, there is a structure in the optic flow, which is related to the direction towards which the bird is flying. But this structure is purely sensory: action is not part of the structure but rather causing the structure (without movement, no optic flow). This is most closely related to Gibson's notion of "invariant structure": some sensory structure that is invariant with respect to some action (Gibson 1986).

4) Action can be involved through proprioception. For example, the position of the robot's head is related to the interaural delay of the sound source, when the head moves. The proprioceptive signal is directly linked to action, rather than to the world. The notion of "sensorimotor contingency" is closest to this case.

5) Action can be seen as an event causing a change in structure or in a sensory signal. For example, a rotation of the head by a given angle changes the sensory structure from one interaural delay to another one. This is an event in the sense that it is transient (an action is not a flow), while the sensory structure is defined on a temporal flow. This can be seen as a mapping from (action, structure) to structure. Note how this is different from the previous case: action cannot be mapped to sensory structure but rather to a change in structure. This is a very important notion that is related to the mathematical notion of *group action*. It corresponds to what Henri Poincaré meant when claiming that the Euclidean structure of space derives from the relationship between movements and sensory inputs (Poincaré 1968). In neuroscience, action defined in this sense corresponds to the notion of "efferent copy", the copy of motor commands that is available for the perceptual system.

4.2. Conditional and unconditional structure

There are two very different types of structure. One is the structure that is normally present in the world, which I will call *unconditional structure*. This is typically the relationship between actions and proprioceptive signals, for example signals that indicate the state of muscles or the relative position of different limbs. The same action normally results in the same change in proprioceptive signals, and the proprioceptive signals do not change unless an action is performed. On the other hand, when a source makes sound, it produces a sensorimotor structure that is *conditional* to the presence and direction of that source: the structure is indeed informative about the source precisely because it is conditional to it and to its properties. It can be said that this is what distinguishes the body from the external world.

I will give two anecdotic examples to help clarify this point. One usually feels her teeth as being part of the body – and in fact, in general, not feeling them at all. But when a tooth is removed and replaced by a dental implant of the same size, it initially feels very present and uncomfortable, as if there were an external body in the mouth. And indeed it is true that there is an external body in the mouth. Yet it occupies the same space as before and it is attached to the mouth as before, it simply has a slightly different shape that can be picked up by the tongue. The feeling is the same when a dental implant is renewed, so it does not have to do with the artificial nature of the tooth. And then after a few days, the implant feels like a normal tooth and it becomes difficult to distinguish the artificial tooth from the natural teeth. From the point of view of sensorimotor structure: before and after the tooth is changed, the structure is unconditional, but there is a transient change in structure at the time when the implant is inserted, which normally means that there is an external body in the mouth. The interesting point here is that this structure is carried by sensors, tactile receptors on the tongue, that are not specifically proprioceptive. They are involved for example in the perception of taste. But depending on whether the structure is conditional or unconditional, it signals elements of the body or of the external world.

Another example is the vestibular system, which is involved in the perception of body balance. In the cochlea, there are receptors that sense head acceleration. The vestibular system also integrates multimodal information coming from motor and visual systems. Acceleration sensors qualify as proprioceptive, in the sense that they are normally only affected by self-generated movements: there is an unconditional sensorimotor structure. Interestingly, when there is a dysfunction of the vestibular system (due to a disease for example), so that this sensorimotor structure is disrupted, one feels nauseous. The standard interpretation of this fact is that nausea is a reflex response of the organism to a dysfunction of the vestibular system that is normally due to the ingestion of toxins: the organism then vomits to get rid of the toxins (Treisman 1977). In other words: if the unconditional sensorimotor structure is disrupted then something must be wrong with the body.

Unconditional structure may also be defined in a statistical sense. For example, two nearby photoreceptors on the retina normally receive a similar amount of light. This simple observation provides topological relationships between receptors. In the same way, two neighboring inner hair cells in the cochlea transduce a similar displacement of the basilar membrane. More generally, a topology on sensors may be defined from the statistical structure of sensor data (specifically, from their correlations).

It is particularly important to characterize the unconditional structure, because conditional structure is defined with respect to it. That is, conditional structure is structure that is normally not observed: this is what makes it informative.

4.3. The syntax of subjective structure

I have noted that structure must be understood in a much broader sense than just a mapping between actions and sensory inputs. This was clearly acknowledged by Gibson, who described the subjective structure of light ("ecological optics") in great detail. I will try to outline the main types and properties of subjective structure.

In the example of the robot, I noted that a sound produced by a source induces a particular structure, defined directly on the sensory inputs: an identity between the signal at one ear and the delayed signal at the other ear. This is an invariant structure in the sensory flow, where change is induced by the mechanical vibration of the sound source. As it is defined directly on the sensory inputs, one may call this structure "first-order". But the direction of the sound source is defined by the relationship between head position and interaural delay, that is, between head position and the first-order sensory structure. In this sense, it is second-order structure. This observation implies that subjective structure has two features: compositionality and hierarchy. Compositionality means that a relationship is defined between two or more constituents, and hierarchy means that the resulting relationship can be a constituent of another relationship (higher-order structure).

Another important feature is that subjective structure is contextual (or at least it can be). For example, consider a horizontal field of grass and a visual system looking at it. There are two textures, each of which qualifies as invariant visual structure: the sky, which is (more or less) uniform, and the grass, which is a statistically uniform visual texture on a perspective structure. Each of these structures is seen only in part of the visual field, for specific gaze angles. Thus there are two distinct structures, conditioned to some action (eye movement) or some sensory input (proprioceptive information about eye position).

Thus, subjective structure is in fact very rich, much richer than a mapping between actions and sensory signals. It is compositional, hierarchical and contextual: one may speak of the "syntax" of subjective structure.

### 4.4. Algebraic properties of structure

Subjective structure is in fact even richer: it can have an algebraic structure. The notion of "algebra" means that we are considering operations. This is precisely what an action is. Here we consider the notion that action is something that changes the sensory structure (case #5 in section 4.1). Let us consider again the robot example. When a source makes a sound, a sensory structure with a particular interaural delay is observed. This structure is changed by a rotation of the head. This can be seen as a mapping from (rotation, delay) to delay, that is, from (action, structure) to structure. In other words, the set of actions acts (in the mathematical sense) on the set of structures. It can be seen that this action has the algebraic structure of a group action. Indeed, for every rotation there is an inverse rotation that brings the structure back to its original state; the combination of two rotations is another rotation. The group is in fact commutative. Note that this group structure exists even if rotations are not specified as angles or even as quantities linearly related to angles.

The existence of this algebraic structure in the subjective structure of the world was indeed noticed by Henri Poincaré, who remarked that movements and their relationships to sensory inputs provide us with the geometrical structure of the world.

Thus, there is a very rich subjective structure that exists independently of prior knowledge on behalf of the perceptual system. In principle, this structure can be captured by a perceptual system (provided appropriate cognitive abilities).

### 4.5. Analogy vs. similarity

In this section, I want to touch on the notion of similarity. A standard notion of similarity is that defined mathematically in a metric space. For example two sensory signals are considered similar if their difference is small. A notion of distance is defined on the space of signals, which provides a measure of similarity of any two signals. But there is another notion of similarity that is used in common language, for example in the statement: the heart is like a pump. Here the similarity is not implied in any metric sense. Rather, it is meant that the heart acts on the blood, a body fluid, in the same way as a pump acts on a fluid. The similarity is not about how the objects (heart and pump) look like in some representational space, but about the way they interact with other things. This is what we call an *analogy*. In his infamous criticism of artificial intelligence, Hubert Dreyfus noted that a great aspect of human intelligence is the faculty of analogy (Dreyfus 1992). This notion is also related to Gibson's notion of affordances (Gibson 1986), which are the ways with which we can interact with things in the world (e.g. the ground affords standing, whereas a large volume of water affords swimming).

A famous question in philosophy of perception was formulated by William Molyneux in the 17th century. Imagine a man who was born blind, and who had learned to distinguish by touch between objects such as a sphere and a cube. If one day the man were given sight, would he be able to distinguish and name these two objects by sight alone? This question has generated considerable philosophical literature, and it is not my aim to answer it here (see (Held et al. 2011) for an empirical study). I simply wish to point out that the question makes sense in terms of subjective physics. In vision, light rays reflected by an object are captured by photoreceptors. By moving the eye, the same photoreceptors capture light rays reflected by different parts of the object. In touch, tactile receptors capture mechanical signals at the interface between fingers and the object. A different part of the object is sampled when the fingers are moved across the object. There is some similarity in the subjective structure of the tactile and visual world: for example, for each movement of the eye, there is an opposite movement that makes sensory signals the same as they were prior to the first movement; the same is true for finger movements. This similarity is not metric: it is rather a set of properties that both subjective structures have in common. I propose to call this type of similarity between subjective structures "analogy".

The same physical object may produce analogous subjective structures in two different modalities. Indeed both touch and vision are about spatial configurations of surfaces. For example, a cube has sharp edges. This means that when the eye is moved across a face of the cube, the visual texture is uniform and then there is a discontinuity, which signals the edge. The same discontinuity occurs when a finger is swept across the face of the cube. In contrast, when the finger moves around a sphere, no such discontinuity occurs, as long as the finger is kept in contact with the surface. The situation is slightly different with vision, since there is also a visual boundary for a sphere, but it is a different type of boundary: when the eye is moved across the surface of a sphere, the visual texture progressively becomes denser as one gets closer to the boundary. Thus, there is a partial analogy between the two subjective structures induced by the presence of the same physical object.

Back to our robot, we may imagine a similar question in which the hearing robot is given visual sensors. Is there anything analogous between the sense of direction obtained through vision and through hearing? The sensors are very different, and there is no such thing as an interaural time difference in vision. However, one thing is identical: the rotation commands also act as a group action on the set of visual properties. Analogies may be useful for the perceptual system, as it may allow the system to learn new sensorimotor contingencies faster.

## 5. Advanced subjective physics of a hearing robot

I will now describe the subjective physics of the hearing robot in a less idealized setting. It will illustrate the concepts I have previously presented in practical cases.

5.1. Two ears

I consider again a robot with two ears, which are mounted on a head. Previously, I described a simplified physical situation in which the signals at the two ears are delayed copies of each other. This is not valid if there is an object between the two ears. The presence of the head produces intensity differences between the ears (it casts an "acoustic shadow"), but it also makes timing differences depend on frequency (G. F. Kuhn 1977). The two signals are no longer delayed copies of each other. The physics of the situation is best described as follows: the source signal S(t) is linearly filtered by two acoustical filters, which depend on sound direction (Fig. 7A). These filters are called head-related transfer functions (HRTF) in the frequency domain, or head-related impulse responses (HRIR) in the temporal domain. The monaural signals $S_L(t)$ and $S_R(t)$ are then described as the convolution of S(t) with the two HRIRs:

$S_L = HRIR_L(\theta) * S$

$S_R = HRIR_R(\theta) * S$

where θ is the sound direction. This is the physical description of the situation. However the subjective physics is quite different, since only the monaural signals $S_L(t)$ and $S_R(t)$ are captured: neither the source signal S(t) nor the set of HRIRs is directly captured. Both pieces of information constitute metaphysical knowledge for the perceptual system. When there were no head between the ears, one could describe one signal as a delayed copy of the other, but this is not possible anymore.

However, there is still some invariant structure in these two signals. In particular, there exist two filters $F_L$ and $F_R$ such that $F_L * S_L = F_R * S_R$. Indeed, this is true with $F_L = HRIR_R(\theta)$ and $F_R = HRIR_L(\theta)$ (Fig. 7B). It is also true with the pair of filters $F_L = U * HRIR_R(\theta)$ and $F_R = U * HRIR_L(\theta)$, for any linear filter U. Therefore, there exists a non-unique pair of filters such that $F_L * S_L = F_R * S_R$. This is a fact of subjective physics that does not include metaphysical knowledge. The perceptual system cannot identify the HRIRs or source signal S(t), but it can identify a particular pair of filters that satisfies the above-mentioned identity. This identity is an invariant, it holds as long as the sound has energy and it does not depend on the source S(t) (at least for broadband sounds). Indeed it is related to spatial location through the identity $F_L * HRIR_L(\theta) = F_R * HRIR_R(\theta)$. This is an example of "invariant structure" in Gibson's terminology.

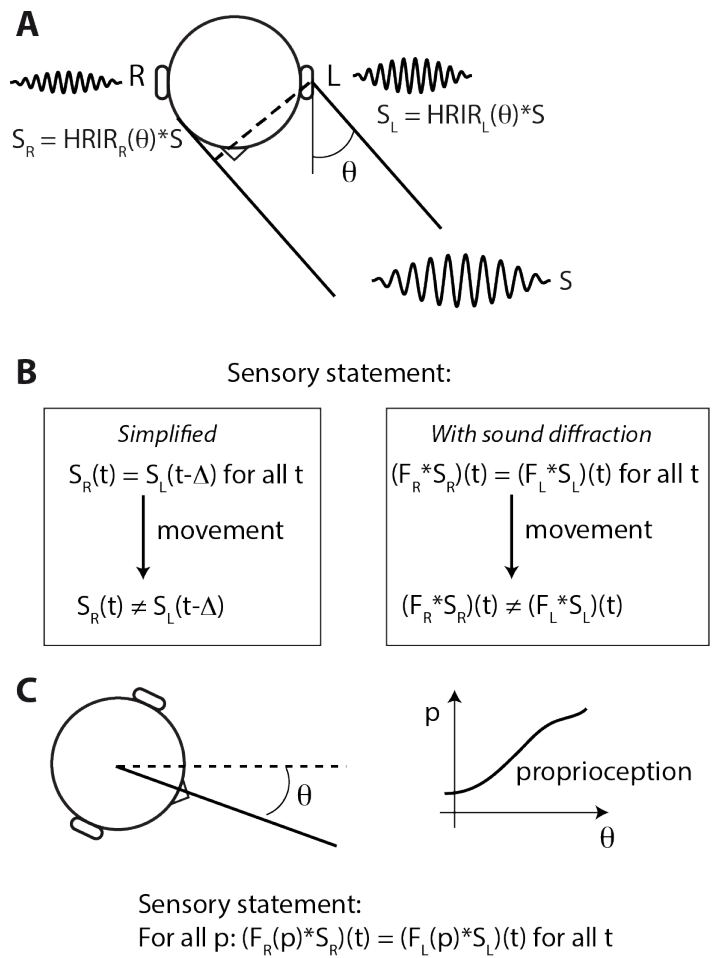

**Figure 7**. Subjective structure of a binaural robot with sound diffraction. A, The signal $S_R$ at the right ear is the result of filtering the source signal S with a directional filter (HRIR for head-related impulse response). B, As in the simplified case without diffraction, the two signals follow a law, which is falsified when a movement is produced. Here $F_L$ and $F_R$ are two filters, for example $F_L$ = $HRIR_R(\theta)$ and $F_R$ = $HRIR_L(\theta)$ (or any filtered version of this filter pair). C, Sound source location is specified by the relationship between a proprioceptive signal p and the sensory law followed by the auditory signals (characterized by a filter pair $F_L(p)$, $F_R(p)$).

The identity is invalidated when the robot turns its head, which makes it a spatial invariant. Without a head, there is a relationship between proprioception (related to the head's angle) and interaural delay, which can be described by a real-valued function. The situation is slightly more complex here: there is a relationship between proprioception and sensory statements of the kind $F_L*S_L = F_R*S_R$. This can be summarized by a mapping from proprioceptive signal p to filter pair ($F_L$, $F_R$) (Fig. 7C). This mapping constitutes the subjective location of the sound source. The effect of actions on the structure is then identical to the idealized case with delays (the subjective structures are analogous in the sense proposed in section 4.5).

This principle has been used in two state-of-the-art sound localization algorithms. One algorithm consists in calculating the convolutions $HRIR_R(\theta)*S_L$ and $HRIR_L(\theta)*S_R$ for all measured pairs of HRIRs, and identifying the pair that maximizes the correlation between the convolved signals (MacDonald 2008). Another algorithm uses a similar technique, but refined in frequency bands

(Durkovic et al. 2011). Note however that the two techniques rely on prior knowledge of the HRIRs.

This viewpoint has also been developed in a spiking neural model of sound localization (D. F. M. Goodman and Brette 2010). There the filters ($F_L$, $F_R$) are assumed to represent the auditory receptive fields of monaural neurons on both sides. The acoustical invariant $F_L*S_L = F_R*S_R$ is then reflected by a synchrony invariant: neurons with filters $F_L$ and $F_R$ fire in synchrony when the sound is presented at a specific location (see (Brette 2012) for a more general framework). Thus a pattern of synchrony is a signature of a particular invariant, and it must then be associated with the corresponding proprioceptive signal. In principle, this model does not require prior knowledge of the HRIRs, provided there is sufficient diversity in the neural filters (see (D. Goodman and Brette 2010) for a simplified learning procedure).

5.2. One ear

I now describe the subjective physics of sound localization with a single ear. The ear is mounted on a head that acts as an acoustic shadow for the ear (Fig. 8A). Consequently, sound intensity at the ear depends on the position of the source relative to the ear. Physically, the relationship between the source signal S(t) and the signal at the ear X(t) is linear: $X(t)=a(\theta).S(t)$, where $a(\theta)$ is the attenuation for sound direction θ. In fact, more rigorously, the attenuation should be described as a filter: $X = a(\theta) * S$. However, I will just consider the simplified setting.

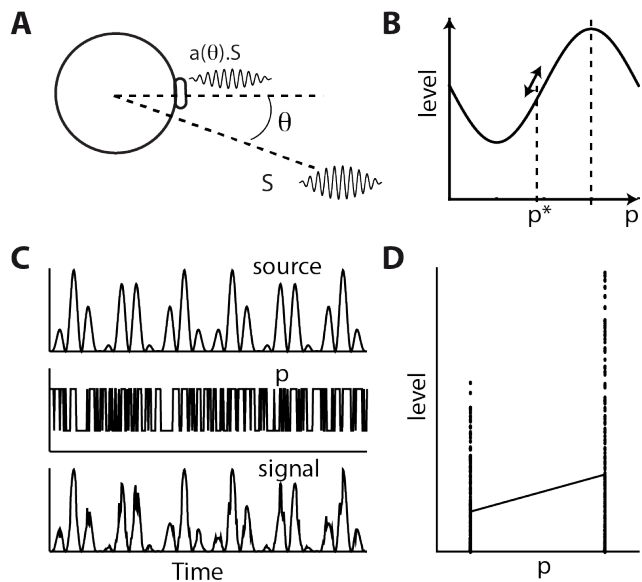

**Figure 8**. Subjective physics of a robot with one ear. A, The attenuation of the source signal at the ear is determined by the angle of the source relative to the ear. B, The level at the ear varies with proprioceptive signal p, which is related to head angle in an unknown way. At position p*, the source is said to be "on the right" of the robot because increasing p ("turning right") increases the level, for a constant source level. C, If the source level is varying (top) while the head is moved (middle), then changes in level at the ear (bottom) cannot be directly attributed to source direction. Here the head was moved alternatively and at random between two positions around p*. D, If movements are independent from the source, as in C (middle), then on average the relative level at the two positions reflects the direction-dependent attenuation shown in B (solid line connects the two averages).

I will address two questions of subjective physics: 1) can there be a non-metaphysical notion of space, given that S(t) is not observed and possibly non-stationary?, 2) what does "in front" mean?

If the source is on the left, then turning the robot's head to the left would make sound level increase (Fig. 8B). The opposite observation can be made if the source is on the right. If the source is right in front, or in the back, then any movement makes sound level decrease. This seems to define three spatial concepts from a subjective viewpoint. The difficulty, of course, is that the source signal S(t) is different before and after the movement. The level at the ear may change because of the movement or because the level of the source has changed (Fig. 8C).

This is a case where the concept of voluntary action is important. In section 3.2, I explained that action is considered voluntary in the sense that it is not caused by sensory signals. On the contrary, action causes changes in sensory signals. The exact opposite would be a reflex (a motor command caused by sensory signals). Suppose the robot moves its head randomly between two positions $\theta-d\theta$ and $\theta+d\theta$ while the sound plays, and the squared signal (for example) is continuously measured: $X^2(t) = a(\theta(t))^2 \cdot S(t)^2$. Then, the expectation of $X^2$ given that the robot's head is at $\theta-d\theta$ is $a(\theta-d\theta)^2 \cdot E[S(t)^2]$ and similarly for the other position. It follows that the relative values of $a(\theta-d\theta)^2$ and $a(\theta+d\theta)^2$ can be observed (Fig. 8D). The same reasoning holds for the entire function $a(\theta)^2$, which can be observed up to a scaling factor. This function defines the subjective location of the sound.

This example was just meant to demonstrate that, because action is voluntary, it is possible for a perceptual system to extract spatial information in a situation that seems intrinsically ambiguous. Note that in contrast with the binaural case, inference is not possible without movement – unless the set of sounds is statistically constrained and statistical inference can be used.

What does "in front" mean? The source is "in front" when the level of the sound is maximal. The source is "in the back" when the level of the sound is minimal. This is a fine definition, assuming the notion of level is known. Specifically, the potentially problematic notion here is "maximal level": this implicitly requires that the value provided by the level sensor positively correlates with sound level. Without this implicit knowledge, an alternative definition can be proposed, by thinking of the moveable ear as an information-seeking device. Consider some omnidirectional background noise in addition to the source signal. The level of the source then acts on the signal-to-noise ratio, and therefore on the intelligibility of the source. Therefore, we may redefine "in front" as the spatial location of maximal intelligibility. Intelligibility requires to also define the content of signals ("what"), not just their spatial location ("where"). For example, if source signals consist of pure tones of various frequencies and levels, intelligibility can be defined as the degree of predictability of the signals, which depends on the signal-to-noise ratio. This provides a definition of "in front" that is independent of the particular way level is transduced by sensors.

Other concepts can be defined, in relationship with the effect of infinitesimal movements:

- "in the axis of the ear" is when small movements have minimal impact on the signal (corresponding to front or back). This definition may be independent of intelligibility.

- "on the left" (resp. "on the right") is when a local movement in the clockwise (resp. anti-clockwise) direction decreases level or intelligibility.

## 6. Subjective physics of light, sound and touch

### 6.1. Hearing vs. seeing

Previously, I have chosen examples taken from auditory perception. Subjective physics applies to hearing, but also to vision and touch. In this section I will try to describe analogies and differences between these sensory modalities. Physically, light is mediated by electromagnetic waves just as sound is mediated by acoustical waves. One can speak of the spectrum of light or of a sound, of light and sound diffraction, etc. Thus from the point of view of physics, there are many similarities. From the point of view of subjective physics, there are many differences, which I will try to outline.

The subjective physics of vision was described in great detail by James Gibson under the name "ecological optics" (Gibson 1986). I will quickly summarize his view here. Illumination sources (the sun) produce light rays that are reflected by objects. More precisely, light is reflected by the surface of objects with the medium (air, or possibly water). What is available for visual perception are surfaces and their properties (color, texture, shape…). Both the illumination sources and the surfaces in the environment are generally persistent. The observer can move, which changes the light rays received by the retina. These changes are highly structured because the surfaces persist, and this structure is informative of the surfaces in the environment. Thus visual subjective structure corresponds to the arrangement and properties of persistent surfaces. Persistence is crucial here, because it allows the observer to use its own movements to learn about the world.

On the other hand, sounds are produced by the mechanical vibration of objects. This means that sounds primarily convey information about volumes rather than surfaces. They depend on the shape but also on the material and internal structure of objects (for example whether the object is full or empty). It also means that the information in sounds is about the source of the waves rather than their interaction with the environment, in contrast with vision. Crucially, contrary to vision, the observer cannot directly interact with sound waves, because a sound happens, it is not persistent. An observer can produce a sound wave, for example by hitting an object, but once the sound is produced there is no possible further interaction with it. The observer cannot move to analyze the structure of acoustic signals. The only available information is in the sound signal itself. In this sense, sounds are events (Casati and Dokic 1998; O'Callaghan 2010). There are of course some properties of sounds that are persistent: precisely the spatial properties (location of the sound source), as we have seen before. But the shape of an object is not specified by sound in the way it is specified by vision. In vision, the way the visual signals change when one moves around an object specifies the shape of the object. When the observer moves around the source of a sound, even a stationary one, the acoustical waves do not change is such a lawful way. This is not to say that the sound contains no information about shape. Indeed the structure of the acoustical signal is related to properties of the sounding object, in particular material and shape (Gaver 1993). For example, the resonant modes are informative of the shape. However, the relationship between this structure and the three-dimensional shape of the object is metaphysical knowledge (if only auditory signals are available).

These observations highlight major differences between vision and hearing from the viewpoint of subjective physics, which go beyond the physical basis of these two senses (light waves and acoustic waves). Vision is the perception of persistent surfaces. Hearing is essentially the perception of mechanical events on volumes.

6.2. Touch

How about touch? As I previously mentioned when discussing Molyneux's problem, tactile perception is sometimes likened to vision in philosophy, by identifying the light ray impinging on a photoreceptor with a finger. This analogy also shows that the subjective structure of touch includes information about the spatial arrangement of surfaces, as vision, and the objects of touch are persistent (even though contact may not be). There are important differences with vision. There is no illumination source. Touch is a proximal sense that requires contact, whereas vision and hearing are distal senses. This means in particular that distance manifests itself in a different way in the subjective structure of touch: distance is defined not by the effect of movements on some continuous property of the sensory signals (level, visual solid angle), but by the movement necessary to make contact with the object. Touch is also about volume, or more precisely about weight, and also about the type of material (soft/hard) in the relationship between hand movements and mechanical signals. Other aspects of active touch have been described by Gibson (GIBSON 1962).

There is also an analogy between sound and touch. I previously mentioned that sounds are about mechanical events on volumes, but this is incomplete. A mechanical event implies an interaction localized at the surface of the resonating object. This can be an impact, or more continuous interactions such as scratching or rolling. Information about the surface is then present in the temporal structure of the sound (Gaver 1993). This is not about the spatial arrangement of the surface, but rather its texture. Such interactions result in mechanical waves that are tactile and auditory, possibly also visual. Therefore in the sensorimotor relationship between finger movement (or eye movement) and mechanical waves (or visual signals), textures produce an analogous subjective structure for hearing, touch and vision.

## 7. The mind-world boundary

The subjective structure of the world is determined by the interface between the perceptual system and the world, that is, by the set of sensors and possible actions on the world. This interface is traditionally located at the physical interface between the body and the external world. However, the exercise of subjective physics could as well be applied to less obvious interfaces, of which I will now give a few examples.

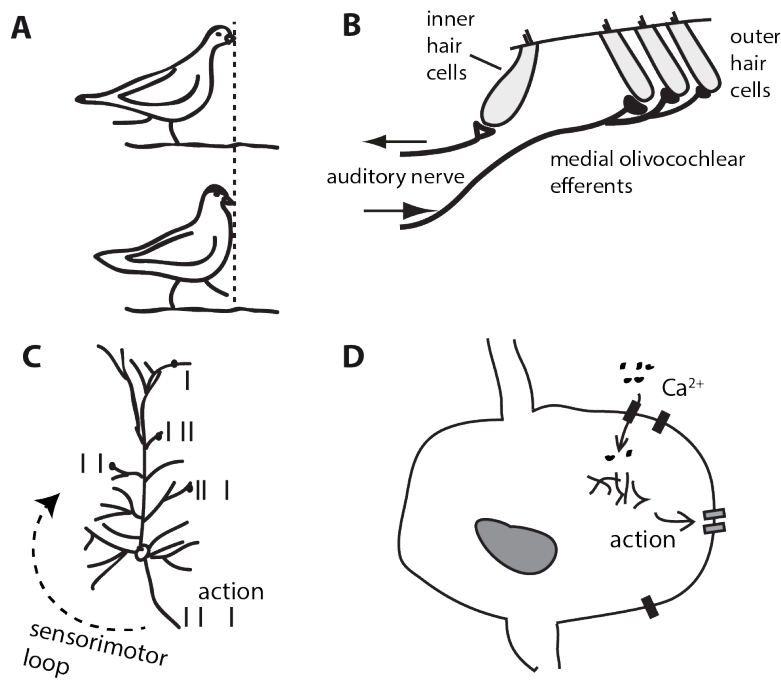

**Figure 9.** The mind-world boundary. A, Some birds move their head as they walk, in a characteristic fashion called "head-bobbing", such that the head is stationary relative to the ground except at discrete moments. It changes the effect of locomotion on the visual field. B, The cochlea can be considered as a sensorimotor system, in which sensors are the afferent fibers capturing the vibration of the basilar membrane via the inner hair cells, and actuators are the efferent fibers influencing that vibration through the outer hair cells (adapted from (Guinan 2006)). C, A neuron can be seen as a perceptual system, where the synapses are its sensors, its axon is its actuator, and all the other neurons together with the outside environment are "the world". D, The same neuron can be seen as a perceptual system of a different kind, by redefining the sensors as molecular signal detectors (e.g. calcium) and actions as expression or regulation of ionic channels and other elements of structure.

7.1. Subsumption architectures

A subsumption architecture is a way to decompose complex behaviors into a set of layers, with each layer taking control ("subsuming") over the underlying layers. It was introduced in the field of behavior-based robotics by Rodney Brooks and colleagues (R.A. Brooks 1986), as a strategy to develop autonomous robots of increasing complexity, but the concept may also apply to biology. For example, there is a reflex that makes the eye follow a slowly moving object appearing in the fovea. As far as subjective physics is concerned, this reflex may be considered as part of the perceptual system (one of the possible actions) or as part of the world, i.e., as something that happens independently of the perceptual system. The subjective structure is different in both cases. If it is considered as being in the world, then motion of an object corresponds not to a displacement of the visual field, but to a change in proprioceptive signals (motion of the eye), together with a change in the background.

In the same way, actions may be considered as direct controls of the physical actuators, or as commands on structures that may issue complex motor sequences, possibly involving control mechanisms, as in servomotors for example. A biological example is found in the locomotion of some birds like pigeons (Fig. 9A). When these animals walk, their head displays a characteristic forward and backward movement called "head-bobbing" (Necker 2007). Each cycle consists of a rapid forward movement of the head (thrust phase) followed by a phase where the head is stable relative to the ground, i.e., moving backward relative to the body so as to compensate for locomotion. It follows that the visual field changes in almost discrete steps (during the thrust phase) when the body moves continuously.

7.2. The cochlea as a sensorimotor system

The cochlea is the organ of hearing. In the cochlea is the basilar membrane, which the acoustic wave received at the ear puts in motion. The stiffness varies from base to apex, and different places along that membrane are maximally sensitive to different frequencies. Thus the acoustic wave elicits a spatiotemporal pattern of vibration. Inner hair cells sit on the basilar membrane and transduce mechanical vibrations into electrical signals, which are in turn converted to spikes in the auditory nerve (Fig. 9B). The nerve then projects to various nuclei in the auditory brainstem. This process is generally depicted as a frequency analysis, as if a bank of band-pass filters were applied to the acoustic wave.

There are also other cells sitting on the basilar membrane, the outer hair cells. They are more numerous and they are thought to actively adjust the "gain" of the cochlea, i.e., to be responsible for the dynamic compression that allows the 100 dB dynamic range of human hearing, and maybe to protect the cochlea against loud sounds. They receive inputs from olivocochlear efferents, with bodies in the superior olivary complex (also in the auditory brainstem), mostly on the contralateral side (Guinan 2006). These are contacted by neurons in the cochlear nucleus, which themselves receive input from inner hair cells (through type I auditory nerve fibers). There are also (ipsilateral) olivocochlear efferent inputs onto inner hair cells. These are less numerous and seem to target the initiation site of type I auditory nerve fibers. I will focus on the most abundant ones, which target the outer hair cells.

Olivocochlear efferents can change the amplitude, but also the phase of the basilar membrane displacement in response to a sound (Cooper and Guinan 2003). It would then be conceivable to examine the subjective physics of the cochlea seen as a sensorimotor system, where sensors are the afferent fibers (capturing the vibration of the basilar membrane), and actuators are the efferent fibers (influencing that vibration through the outer hair cells). There is then information about a sound in the relationship that it induces between the actions on outer hair cells and the inputs from the inner hair cells.

7.3. The neuron as a perceptual system

Electrical communication between neurons is directional: action potentials propagate along the axon of a neuron, from the initial segment near the cell body to the synaptic terminals, where they produce electrical changes in other neurons. Therefore one could consider the synapses of a neuron as its sensors, its axon as its actuator, and all the other neurons together with the outside environment as "the world" (Fig. 9C). For the neuron seen as a perceptual system, spikes are actions that it produces on the world. The subjective physics of this system describes the relationship between spikes and their effects on the neuron's inputs. Somehow, this is what the neuron can "know" about the outside world.

The neuron could also be seen as a perceptual system of a different kind, by redefining the set of sensors and actions (Fig. 9D). Calcium is a universal signal in cells in general, and in neurons in particular (Berridge, Lipp, and Bootman 2000). It triggers a large number of processes such as the expression of ionic channels or morphological changes. The cell could then be seen as a perceptual system with these processes as the actions it can take, and calcium sensors as its sensors; sensors could also be voltage-activated calcium channels, for example. The subjective physics of this system describes the relationship between plastic changes (synaptic changes, ionic channel regulation) and their effect on calcium signals and other signals captured by the cell.

## 8. Discussion

I have proposed to define subjective physics as the field of study that analyzes the structure of the sensory and sensorimotor relationships (laws) that are available to an organism embedded in an environment, given a set of sensors and actuators, without metaphysical knowledge about the world. I will first summarize the main concepts in this text, and then discuss the relevance of subjective physics to various fields.

### 8.1. Summary

Subjective physics describes a set of laws, in the same sense as physics describes a set of laws. The difference is that these laws are seen from the perspective of the perceptual system, the analog of the scientist. The perceptual system derives its knowledge of the world from its sensors and the actions it can produce – the "interface" with the world, in the same way as a scientist has access to measurements and can make experiments. This knowledge is a set of laws, which can take two forms: *weak knowledge*, the kind that has a predictive value about future sensory inputs; *strong knowledge,* the kind that applies to the effect of actions on sensory inputs. Any other form of knowledge is called *metaphysical knowledge*, in analog with Popper's demarcation criterion in philosophy of science. It encompasses all knowledge that an external observer might have, but that does not derive from the relationships between actions and sensors.

This set of laws is called the *subjective structure of the world*. It can be progressively analyzed at different depths of *ecological reduction*, which I defined as the suspension of prior knowledge about the world on behalf of the perceptual system: at the first level, we may consider that sensor data is specified in external terms (acoustical pressure); at the second level, we may consider that sensor data is specified in internal terms (transduced quantity); at the third level, we may consider that sensors are inhomogeneous.

A sensor is something that is modified by the world, and whose modification can be picked up by the perceptual system. An action is something that the perceptual system can do, which produces modifications in the world and can ultimately affect sensors. In subjective physics, actions are considered *voluntary*, in the sense that any action can potentially be taken. A consequence is the existence of causality in the sensorimotor structure: causality can be distinguished from correlation because action can be taken or not taken, independently of sensory inputs. Action can be involved in the subjective structure in five different ways: not at all (sensory structure); structure conditioned to the absence of action; action as a cause of the sensory flow (without which there is no structure); through proprioception (sensorimotor contingencies); as an event causing a change in structure or in a sensory signal.

Subjective structure can be unconditional: a set of laws that are always satisfied. Or it can be conditional, that is, not normally observed (e.g. induced by the presence of an object). Conditional structure is therefore defined with respect to unconditional structure (the reference). In contrast with classical notions of information (Shannon), which are unstructured, subjective structure is compositional, hierarchical and contextual: one may speak of the "syntax" of subjective structure. Actions, seen as operations on the structure, define algebraic properties (e.g group action). This rich notion of structure provides a notion of analogy that is different from similarity between metric spaces: subjective structures are said to be analogous when they have a shared set of properties.

I have shown how to apply these concepts to two fictional examples involving a hearing robot. Applied to humans, subjective physics of light, sound and touch have analogies and differences, which go beyond their physical substrate (different types of waves). For example, vision is the perception of persistent surfaces, while hearing is essentially the perception of mechanical events on volumes.

The subjective structure of the world is determined by the interface between the perceptual system and the world, that is, by the set of sensors and possible actions on the world. As far as the exercise of subjective physics is concerned, any arbitrary interface can be defined. For example, the cochlea or a neuron can be seen as perceptual systems. In the first case, the "world" is the basilar membrane; in the second case, it is the rest of the brain and the external environment.

I will now discuss the relevance of subjective physics for various fields.

8.2. Subjective physics and computational neuroscience

I started this essay with a quote from David Marr, a major figure in computational neuroscience: *"If we are capable of knowing what is where in the world, our brains must somehow be capable of representing this information."* (Marr 1982). For a perceptual system, "knowing what is where in the world" corresponds to what Marr called the "computational level", what the system is supposed to do.

This text is largely an attempt to redefine the computational level of perceptual systems in a way that does not rely on objective descriptions made by an external observer. Instead, the computational level is described from the perspective of a perceiver embedded in its environment. This viewpoint departs from traditional approaches in computational neuroscience of perception, which follow the information-processing paradigm: the system is given some sensory inputs, for example a pair of acoustic signals, and it processes them into an output, which could be the estimated angle of the sound source. The output is given a meaning by the external observer, but for the perceptual system itself, it is nothing else than a real number. In this paradigm, the task of the system is to achieve a particular transformation between inputs and outputs.

This traditional approach implicitly assumes that the system has metaphysical knowledge about the world, in the sense that I proposed. If we do not want to make this assumption, then the traditional approach is not satisfying as a way to understand perceptual systems. In this case, the task of the perceptual system is radically different: it is not to achieve a particular transformation anymore, but to grasp the laws of subjective physics. The relevant analogy is not that of a machine applying a sequence of operations to inputs so as to produce an output, but rather that of a scientist who can make measurements and experiments, and who draws conclusions in the form of laws.

How should neural models of perceptual systems look like, in the framework of subjective physics? In subjective physics, the basic element of perception is a law that sensory signals follow (in a broad sense, including proprioceptive signals), or "invariant structure" in the terms of Gibson. Therefore, one basic challenge for neural models of perceptual systems is to identify and respond to laws that unfold in time. There is some work in the field that is relevant to this theme. One relevant line of work is a set of learning algorithms based on the "slowness principle" (Földiák 1991; Mitchison 1991; Becker and Plumbley 1996; Stone 1996; Wiskott and Sejnowski 2002). The idea is that characteristics of the world, for example the location of a sound source, vary more slowly than sensory signals that are caused by them (e.g. acoustical signals), another way to express the idea that laws are invariant while the constituents of the laws are variable. The learning algorithms then consist in projecting the signal space into another space where projected signals vary as slowly as possible. These projected signals are then expected to be characteristics of laws.

Another relevant line of work is the idea that synchrony is a temporal invariant, and if spike trains are caused by sensory signals, then a particular pattern of synchrony in neural population reflects the occurrence of a particular sensory law (Brette 2012; D. F. M. Goodman and Brette 2010). I defined the "synchrony receptive field" as the set of stimuli that elicit synchronous firing in a given group of neurons: it corresponds to a temporal invariant or law. A neuron that detects coincidences between these neurons then spikes when the stimulus follows that law.

These approaches are promising but they only address a small subset of the concepts developed here. For example, computationally speaking, detecting that sensory signals follow a particular law is not the same as predicting the future of these signals, or predicting the effect of an action on future sensory signals. Perhaps more importantly, I have shown that the subjective structure of the world has syntax: it is compositional, hierarchical and contextual; actions add an algebraic structure. These properties are ignored in standard representational theories in neuroscience because neural representations are intrinsically unstructured. In the concept of "neural assemblies", which is the mainstream assumption about how things we perceive are represented in the brain, any given object is represented by the firing of a given assembly of neurons. Therefore, the structure of such neural representations is the structure of subsets of a fixed set of elements: a neural assembly is a "bag of neurons", in the same way that search engines analyze the content of a web page as a "bag of words" with no relationships between the words. This weakness has in fact been observed many times in the past in the context of the "binding problem" (von der Malsburg 1999): when two objects are present, they are represented by a merged assembly in which the identity of the two sub-assemblies is lost (the so-called "superposition catastrophe"). The identification of this weakness led to alternative propositions, such as using time as a signature of represented objects ("binding by synchrony" (Singer 1999)). Such representations are richer (in particular compositional and hierarchical), but still not as rich as subjective structure: sets of features can be represented, but not relations between features, apart from belonging to the same set. For example, the statement "action A changes sensory structure B to C" describes a relationship between B and C, labeled by A. This does not fit the neural assembly framework, even augmented with binding with synchrony. Other authors have proposed that taking into account the order of activation of neurons provides a syntax to neural firing, which may be a way to address this problem (Buzsáki 2010).

Finally, even the most sophisticated representational theories still pose a problem to the viewpoint of subjective physics, because they leave the interpretation of the neural representations to some unspecified external observer. A simple way to solve this problem is to not have an interpretation stage. That is, instead of considering neural models that "represent" the world, one can consider neural models that produce actions in the world; in other words, models that are *autonomous*. This theme has been developed in behavioral robotics (Rodney A. Brooks 1991) and in enactive philosophical theories of the mind (Thompson and Stapleton 2009), but not very much in computational neuroscience. Yet it seems almost unavoidable that neural models of perception developed in the framework of subjective physics must be autonomous: it is a natural consequence of the attempt of removing metaphysical knowledge on behalf of the perceptual system.

## 8.3. Subjective physics and philosophy of science

In this text, I have made a number of analogies with philosophy of science. This was not accidental. Starting from the observation that what a scientist can know about the world derives from her senses and the actions she can take in the world (experiments, possibly involving measurement devices), philosophy of science asks questions such as: what is knowledge? how can it be acquired? how can we distinguish between contradicting theories?

So there is in fact a formal analogy between these questions and corresponding questions in subjective physics, where sensors are the observation devices and actions are the kind of experiments the perceptual system can make. This is an interesting analogy, because concepts developed in philosophy of science are directly relevant to subjective physics. I will discuss a few of them.

The first remark is that science, like subjective physics, takes the form of universal statements or laws. A law is more than a collection of observations: it says something about observations that have not been made yet – this is what makes science useful. But how are laws formed? The naive view, classical inductivism, consists in collecting a large number of observations and generalizing from them. For example, one notes that all men she has seen so far have two legs, and concludes that all men have two legs. Unfortunately, inductivism cannot produce universal laws with certainty. It is well possible that one day you might see a man with only one leg. The problem is that there are always an infinite number of universal statements that are consistent with any finite set of observations.

Therefore, inductivism cannot guide the development of knowledge. Karl Popper, possibly the most influential philosopher of science of the twentieth century, proposed to address this problem with the notion of *falsifiability* (Popper 1959). What distinguishes a scientific statement from a metaphysical statement is that it can be *disproved* by an experiment. For example, "all men have two legs" is a scientific statement, because the theory could be disproved by observing a man with one leg. But "there is a God" is not a scientific statement. For any practical purpose, only scientific statements are useful, since a metaphysical statement can have no predictable impact on any of our experience, otherwise this would produce a test of that statement.

These concepts can be directly applied to subjective physics: knowledge about the world takes the form of laws; only those laws that could potentially be falsified in the future are useful from the perspective of the perceptual system; other kinds of laws can be considered "metaphysical".

I will not do an exhaustive review of philosophy of science, but I wish to point out that there are many other concepts that are directly relevant to subjective physics. I will mention a few of them. While Popper explains what a scientific statement is and how it can be tested, he does not explain the difficult part, which is how a scientific statement is made in the first place. From a logical point of view, there are an infinite number of possibilities given a finite set of observations. Which one should be chosen? A popular heuristic is "Occam's razor", i.e., the idea that among competing hypotheses, the most parsimonious one should be preferred. This is a well-known concept in statistical learning theory, related to the problem of "overfitting": a simple law is more likely to generalize well than a complex one. But choosing the simple theory also means choosing a theory that is *not* consistent with observations. And indeed Post-Popperian philosophers and historians of science have argued that a scientific theory is not only a theory that *can* be falsified, it is a theory that *is* actually falsified (Lakatos et al. 1978; T. S. Kuhn 1962; Feyerabend 2010). This is made possible by treating falsifications of theories as anomalies, which can be explained by auxiliary hypotheses. For subjective physics, these concepts mean that consistency with observations is not the only criterion that a perceptual system should use to make laws – simplicity could be an additional one, or analogy with other laws (in the sense that I previously defined).

The most radical critics of Popper have made a remark that is highly relevant to subjective physics (Feyerabend 2010; T. S. Kuhn 1962). There is no such thing as an objective observation independent of any scientific theory. Observations are produced by scientists themselves, in the context of the theory they currently favor. Theories are not derived from observations, but rather there is a circular relationship between them. It is in fact implicit in Popper's exposition of falsifiability: a scientific theory should suggest a critical experiment that may falsify it; therefore, it drives future observations. In the same way, observations made by a perceptual system are not independent of that perceptual system. They depend on the actions taken by the system. This suggests that the formation of laws should not be thought of as a process of fitting a curve to a given set of data points, but rather as an active process in which observations are made so as to help the formation of laws. This relates to the concept of active learning (Cohn, Ghahramani, and Jordan 1996). Note that the fact that the relationship between observations and theories is circular does not imply that theories are arbitrary, since observations still depend on the world. But it does imply that the formation of theories (laws) depends on the history of the process.

To end this section, I may venture to propose that conversely, subjective physics may perhaps provide a simple conceptual framework in which to develop concepts of philosophy of science, instead of the more traditional framework of history of science.

8.4. Subjective physics and psychological theories of perception

There is a strong relationship between subjective physics and psychological theories of perception, mostly two of them: Gibson's ecological approach to perception (Gibson 1986), and O'Regan's sensorimotor theory of perception (O'Regan 2011). According to Gibson, the "information to be perceived" is the invariant structure in the sensory or sensorimotor flow, i.e., the subjective structure. Gibson proposed informal descriptions of that structure mainly in vision, with the concept of the "optical array" (Gibson 1972). He also wrote a less detailed account about active touch (GIBSON 1962), and his approach was applied to sounds by Gaver (Gaver 1993).

A major Gibsonian theme is the notion of "affordances". Gibson considered that what we perceive in the world is affordances, a term he coined to designate the possibilities of action that an object in the world allows us to do. For example, a car is something that we can drive, an opening in a cave is something we can go through. According to Gibson, the world is perceived through the actions we can do in it. It is these affordances that produce meaning for a particular organism, and the same object in the world can mean completely different things for organisms that act differently. Subjective physics may provide a framework to study affordances.

According to the sensorimotor theory of perception (O'Regan 2011), what we perceive is the expected effect of our own actions on sensory signals. Subjective physics describes this expected effect. There are a number of differences with Gibson. One is that it grants an important role for inference in perception, which Gibson largely downplayed (he considered on the contrary that perception is direct). But what is inferred is a law (or sensorimotor contingency), not an objective "thing" in the world. A second important difference is that it refines the notion of invariant structure by acknowledging that the signals that the brain receives are only indirectly related to physical signals (light). Finally, in terms of subjective physics, the theory also proposes that the phenomenological structure of conscious perception reflects the subjective structure of the world.

An application of the sensorimotor theory of perception is sensory substitution, for example presenting the image of a camera or a sound through a tactile device (Kaczmarek et al. 1991). The theory predicts that sounds or images can be perceived through sensors that are different from those normally associated with the corresponding perceptual modality, provided that sensorimotor contingencies are preserved. What this text suggests is that sensory signals should be presented to the sensors in such a way that the substituted subjective structure is analogous to the original subjective structure, in the sense that I proposed in section 4.5 (that the two structures have properties in common, e.g. algebraic properties).

8.5. Subjective physics, robotics and neuromorphic engineering

I will end this text on a short discussion of the relevance of subjective physics for robotics and neuromorphic engineering.

In robotics, the theme of subjective physics is connected to theories of embodiment, the idea that the body and its interaction with the environment are parts of the cognitive system. The traditional approach in robotics (and more generally in artificial intelligence) is to consider perception as a separate module whose function is to produce an objective representation of the world based on sensory inputs; another module takes decision and actions on the basis of that representation. There are alternative ideas in robotics, in which the robot learns to use its body and sensors, considered as an unknown "envelope", with a general-purpose algorithm, the "kernel" (Kaplan and Oudeyer 2011). In terms of subjective physics, the envelope is the interface with the world and the kernel is the perceptual system. Thus subjective physics describes the structure of the world experienced by the robot through its envelope, which is to be discovered by the robot's kernel.

Some approaches in neuromorphic engineering use low-power analog circuits to model neurons (Indiveri et al. 2011). There are two motivations: to reproduce the computational abilities of biological neural networks, and to develop electronic devices that consume little power. Using low-power components comes at price: there is some tolerance in the properties of the electronic components, which make them partly unknown to the system. This issue corresponds to the notion that the relationship between physical stimuli and the activation of sensors is metaphysical knowledge for the system. Thus neuromorphic hardware must be designed to work in the absence of metaphysical knowledge of component properties (to some extent). Subjective physics describes what can still be known by the system under these constraints, and may suggest ways to deal with this issue, such as statistical homogenization or including the unknown (but fixed) properties in the subjective structure.

This discussion has sketched a number of perspectives for the development of subjective physics along two main lines. One is to develop the description of subjective physics for human, animal and artificial sensory systems. Another one is to develop the modeling of perceptual systems in the framework of subjective physics, an exciting challenge for computational neuroscience.

**Glossary**

- *Action*: something that the perceptual system can do, which produces modifications in the world. In turn, these modifications can affect sensors.
- *Ecological reduction*: the suspension of any prior knowledge about the world on behalf of the perceptual system, which precedes the analysis of the subjective structure of the world.
- *Eidetic variation*: varying the world, the interface and the constraints on the perceptual system, so as to reveal their relationships with the subjective structure of the world.
- *Interface*: a set of sensors and possible actions on the world.
- *Metaphysical knowledge*: statements about the world that cannot be falsified with the available set of sensors and actuators.
- *Perceptual system*: the system that captures sensor data and performs actions on the world.

- *Sensor*: something that is modified by the world, and whose modification can be picked up by the perceptual system.
- *Sensor homogeneity*: the assumption that sensors have the same properties, for example that different microphones provide values that have the same relationship to acoustical pressure.
- *Statistical homogenization*: a procedure by which sensor properties are made homogeneous, by tuning sensors so that their signals have identical statistics.
- *Subjective physics*: the field of study that analyzes the structure of the sensory and sensorimotor relationships that are available to an organism embedded in an environment, given a set of sensors and possible actions.
- *Subjective structure of the world*: the structure of the world, as it can be discovered by the perceptual system without a priori knowledge. It is made of the laws followed by sensor data and the relationships between actions and sensor data.
- *World*: the environment in which the perceptual system operates.